\documentclass[journal]{IEEEtran}
\usepackage{cite,url,subfigure,epsfig,graphicx}
\usepackage{tipa}
\usepackage{makecell}
\usepackage{multirow}
\usepackage{bbm}
\usepackage{mathrsfs}
\usepackage{array}
\usepackage{amsfonts}
\usepackage{cite,url,subfigure,epsfig,graphicx}
\usepackage{amssymb,amsmath}
\usepackage{indentfirst}
\usepackage{pifont}
\usepackage{algorithmic}
\usepackage{algorithm}
\usepackage{cases}
\usepackage{soul}
\usepackage{booktabs}
\usepackage[table]{xcolor}
\usepackage{graphicx}
\IEEEoverridecommandlockouts
\makeatletter
\usepackage{times,verbatim,amsfonts,amsmath,color}
\usepackage{geometry}
\begin{document}

\title{Blockchain-Envisioned UAV-Aided Disaster Relief Networks: Challenges and Solutions}

\author{
\IEEEauthorblockN{Yuntao~Wang\IEEEauthorrefmark{2}, Qinnan~Hu\IEEEauthorrefmark{2}, Zhendong~Li\IEEEauthorrefmark{3}, Zhou~Su\IEEEauthorrefmark{2}\IEEEauthorrefmark{1}, Ruidong~Li\IEEEauthorrefmark{4}, Xiang~Zou\IEEEauthorrefmark{5}, and Jian~Zhou\IEEEauthorrefmark{9}}\\
\IEEEauthorblockA{
\IEEEauthorrefmark{2}School of Cyber Science and Engineering, Xi'an Jiaotong University, China\\
\IEEEauthorrefmark{3}School of Information and Communication Engineering, Xi'an Jiaotong University, China\\
\IEEEauthorrefmark{4}Department of Electrical and Computer Engineering, Kanazawa University, Japan\\
\IEEEauthorrefmark{5}The Second Research Institute of Civil Aviation Administration of China, Chengdu, China\\
\IEEEauthorrefmark{9}China Mobile Chengdu Institute of Research and Development, China\\
\IEEEauthorrefmark{1}Corresponding author: zhousu@ieee.org
}
\thanks{This work has been accepted by \emph{\textbf{IEEE Communications Magazine} in August 2024}.}}

\maketitle

\begin{abstract}
Natural or man-made disasters pose significant challenges for delivering critical relief to affected populations due to disruptions in critical infrastructures and logistics networks.
Unmanned aerial vehicles (UAVs)-aided disaster relief networks (UDRNs) leverage UAVs to assist existing ground relief networks by swiftly assessing affected areas and timely delivering lifesaving supplies. To meet the growing demands for collaborative, trust-free, and transparent UDRN services, blockchain-based UDRNs emerge as a promising approach through immutable ledgers and distributed smart contracts.
However, several efficiency and security challenges hinder the deployment of blockchain-based UDRNs, including the lack of cooperation between smart contracts, lack of dynamic audit for smart contract vulnerabilities, and low forensics robustness against transaction malleability
attacks. Towards efficient and secure blockchain-based UDRNs, this paper presents potential solutions: (i) a series of collaborative smart contracts for coordinated relief management, (ii) a dynamic contract audit mechanism to prevent known/unknown contract vulnerabilities; and (iii) a robust transaction forensics strategy with on/off-chain cooperation to resist transaction malleability attacks.
Our prototype implementation and experimental results demonstrate the feasibility and effectiveness of our approach. Lastly, we outline key open research issues crucial to advancing this emerging field.
\end{abstract}

\IEEEpeerreviewmaketitle

\section{Introduction}
\IEEEPARstart{N}{atural} disasters such as earthquakes, hurricanes, floods, and wildfires often disrupt critical infrastructures and logistics networks, making it difficult to deliver essential supplies to affected populations \cite{rescuechain}. Traditional relief supply management relies on ground transportation, which may be hampered by damaged roads, limited access, or overwhelmed local infrastructure. In such situations, UAVs play a crucial role in assisting existing ground relief networks \cite{MatraciaCM2021}. Particularly, UAVs can be immediately deployed to remote or inaccessible areas for rapid disaster assessment and timely delivery of relief supplies to affected areas \cite{sardo}.
Additionally, UAVs have potential to establish feasible communication links between relief coordinators and equipment~\cite{lvbs}.

UAV-aided disaster relief networks (UDRNs) \cite{MatraciaCM2021,lvbs} leverage a fleet of UAVs equipped with payload capabilities, sensors, and communication modules to enhance relief supply management in disaster-stricken areas. For instance, UAVs can carry real-time video and imaging systems, thermal sensors, or gas detectors to support search and rescue operations and damage assessments. UAVs can also transport essential items such as food, water, and medical equipment promptly, thereby saving lives and reducing losses. Moreover, artificial intelligence (AI) algorithms enable UAVs to autonomously optimize flight paths, adapt to changing conditions, and improve overall operational efficiency.

Despite the numerous benefits, there are increasing trust issues and transparency concerns in collaborative disaster response and relief management in UDRNs.
Trust is crucial for effective coordination and disaster relief management \cite{WangTNSM2024}, particularly among diverse stakeholders such as UAV operators, emergency communication vehicles, government agencies, non-profit organizations (NPOs), and humanitarian groups. A lack of transparency and accountability can erode trust, leading to doubts about the fairness of relief distribution and potential mismanagement of resources.
Efficient coordination among stakeholders is essential for timely disaster response and optimal relief resource allocation. Inadequate coordination among stakeholders, including information-sharing gaps and overlapping efforts, can result in redundant or insufficient allocation of relief materials.

The emerging blockchain technology \cite{10004889} opens up new possibilities for trust-free, transparent, and accountable disaster management in UDRNs. With its decentralized nature, blockchain allows relief organizations to efficiently manage and track the supply chain, ensuring the timely delivery of relief supplies. Additionally, smart contracts can automate resource allocation and streamline relief operations.
In the literature, there are increasing works exploring blockchain-based approaches for secure and efficient UDRNs, including lightweight blockchain design \cite{lvbs}, energy-efficient consensus \cite{rescuechain}, decentralized charity donations management \cite{10004889}, trusted aerial-ground networking \cite{WangTNSM2024}, secure offline transactions \cite{XingTVT2022}, secure knowledge sharing in disasters \cite{PauuIoTJ2023}.

However, blockchain-based UDRNs face the following new challenges {toward secure and efficient relief management}. {First,} while smart contracts automate UDRN transactions efficiently, they often operate in isolation \cite{10004889}, leading to fragmented disaster relief efforts. {Second, the logic dependence of collaborative contracts further complicates the detection of security vulnerabilities.} Traditional static audit methods \cite{securify,oyente} are limited to pre-deployment audit, while existing dynamic methods \cite{evil-under-the-sun,TXSPECTOR} are costly and constrained by EVM bytecode size. {Third, collaborative contracts are also vulnerable to transaction malleability attacks \cite{Trans_Unmalleable}, which can} compromise the integrity and traceability of relief service records in blockchain-based UDRNs. 

\begin{figure*}[!t]
\setlength{\abovecaptionskip}{-0.03cm}\vspace{-3mm}
\centering
  \includegraphics[width=0.75\linewidth]{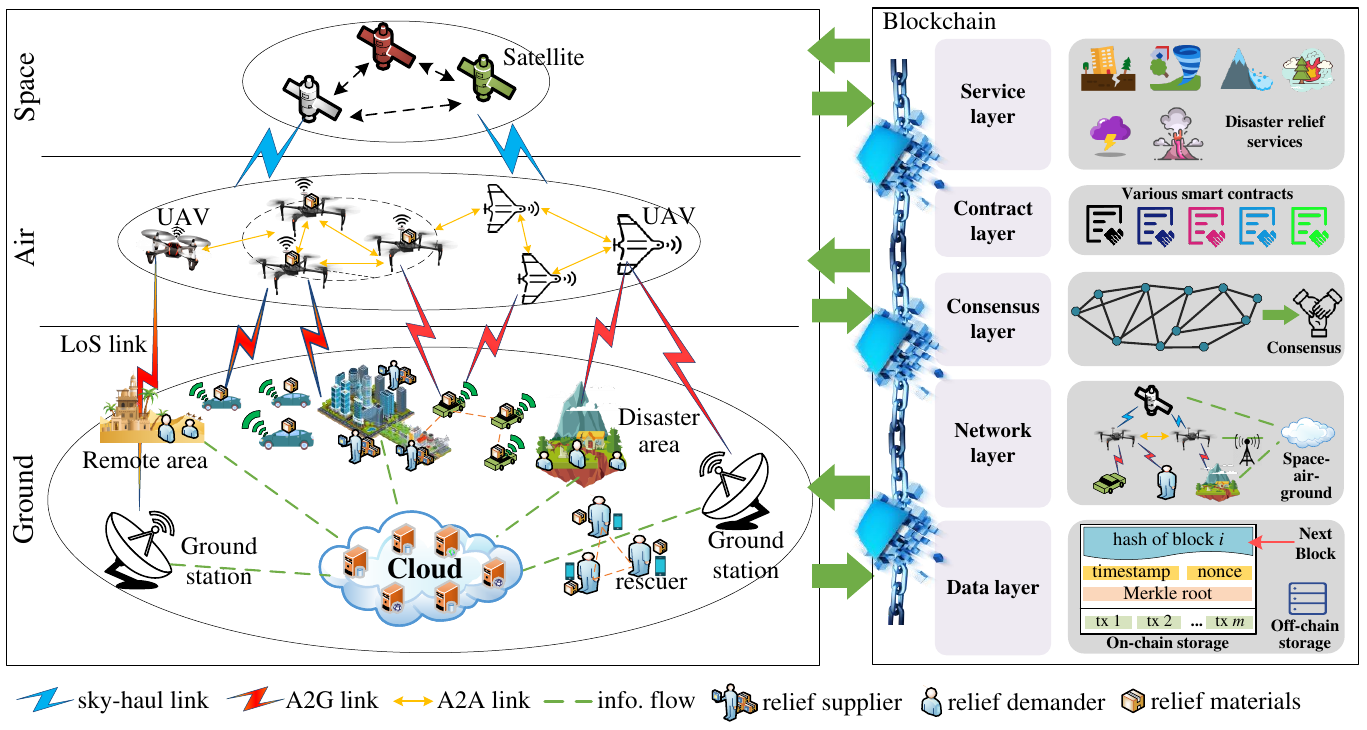}
  \caption{Overview of UAV-aided disaster relief networks (UDRNs) and the general architecture of blockchain-envisioned UDRNs.}\label{fig:1}\vspace{-3mm}
\end{figure*}

To address these challenges, this paper proposes a novel blockchain and smart contract design for secure and efficient relief management in UDRNs. Specifically, we present a general blockchain-oriented architecture for UDRNs that incorporates space, air, and ground layers. Subsequently, we optimize and secure the blockchain-based UDRN system through three key solutions: (i) a series of collaborative smart contracts for automated and coordinated relief management, (ii) a dynamic contract audit mechanism to prevent potential contract vulnerabilities, and (iii) a robust transaction forensics strategy with on/off-chain cooperation to resist transaction malleability attacks.
Through a prototype implementation, experimental results validate that the proposed scheme outperforms existing representatives in terms of transaction retrieval latency and vulnerability detection rate.

The remainder of this paper is organized as below. Section II gives the background of UDRNs and the general architecture of blockchain-based UDRNs.
Section III discusses the key challenges in designing secure and efficient blockchain-based UDRNs. Section IV presents the potential solutions to these challenges.
The prototype design and performance evaluation are given in Section~V. Section~VI discusses open research issues. Finally, Section~VII concludes this work.

\section{Blockchain-Envisioned UAV-Aided Disaster Relief Networks}
\subsection{UAV-Aided Disaster Relief Networks (UDRNs)}
As illustrated in the left part of Fig.~1, a typical scenario of UDRN incorporates three layers: space, air, and ground. It mainly includes the following entities. 

\emph{Satellites} provide reliable and wide-coverage communication services for large-scale rescuers and relief equipment in disaster and remote sites. UAVs can communicate with satellites via sky-haul links. Satellite communications can generally be classified into two types: \emph{broadband} and \emph{narrowband}.

\emph{UAVs} serve as airborne platforms capable of carrying and delivering relief supplies to remote and hard-to-reach locations affected by disasters \cite{MatraciaCM2021}. They can also be equipped with various sensors to gather real-time data for damage assessment. Typically, UAVs play the following roles in disasters.
\begin{itemize}
    \item \emph{Rapid delivery:} UAVs enable swift transportation of essential relief items, such as medical supplies and food to areas inaccessible by traditional means.
    \item \emph{Communication relay:} Via air-to-ground (A2G) and air-to-air (A2A) links \cite{sardo}, UAVs can act as communication relays in areas where terrestrial infrastructure has been disrupted, enabling line-of-sight (LoS) connectivity for relief teams.
\end{itemize}

\emph{Ground vehicles} complement UAVs by providing additional means of transportation and distribution of relief supplies on land. They typically serve the following roles in disaster sites.
\begin{itemize}
    \item \emph{Last-mile delivery:} Ground vehicles transport relief items from central warehouses or distribution centers to locations that are inaccessible to UAVs due to distance or regulatory constraints. \item \emph{On-site support:} Ground vehicles facilitate the movement of relief teams, allowing them to reach affected areas quickly to assess needs and provide assistance.
\end{itemize}

\emph{Ground station (GS)} coordinates UAVs and rescue vehicles at disaster sites and serves as an edge node. It uses real-time data from UAVs, vehicles, and other sources to assess the disaster situation, prioritize response efforts, and allocate resources effectively. The GS can be deployed as either a \emph{fixed} station or a \emph{mobile} emergency communication vehicle \cite{lvbs}.

\emph{Relief cloud center} acts as the central hub overseeing and coordinating all activities in UDRNs \cite{10004889}. It facilitates collaboration among relief suppliers, demanders, and transportation assets (i.e., UAVs and rescue vehicles) to optimize the distribution of relief supplies. It also serves as the communication nexus, ensuring seamless information flow between all components in UDRNs.

\emph{Relief demanders} are individuals, communities, or organizations directly affected by the disaster and in need of assistance. They communicate their needs to the GS or the cloud, specifying the type and quantity of relief materials required at their respective locations. Once the GS processes their requests, UAVs or ground vehicles are dispatched to deliver the necessary supplies.

\emph{Relief suppliers} include governments, NPOs, companies, and local community groups. They collaborate with GSs and the cloud to guarantee that appropriate types and quantities of relief supplies are dispatched to meet the need of affected areas.
\vspace{-3mm}
\subsection{Architecture of Blockchain-Envisioned UDRNs}
In our consortium blockchain-envisioned UDRNs, {there are two kinds of entities: \textit{full nodes} and \textit{light nodes}.} Entities with sufficient computing and storage capacities can serve as full nodes, storing the complete history of the blockchain. In contrast, entities with limited computing and storage capacities {(e.g., UAVs)} act as light nodes, storing only block headers and receiving blockchain services from nearby full nodes.
As illustrated in the right part of Fig.~1, blockchain-envisioned UDRNs typically comprise the following five layers:
\begin{itemize}
  \item \emph{Data layer.} This layer collects and verifies disaster-related information from various sources (including UAVs, ground vehicles, IoT devices, and satellite imagery), to improve data accuracy and ensure data authenticity. The collected multi-source data is collaboratively stored in decentralized and hash-chained blocks, as well as the off-chain storage. 
  \item \emph{Network layer.} The network layer efficiently propagates disaster-related data from diverse sources to all involved nodes in UDRNs through space-air-ground integrated networking, enabling real-time data exchange and information synchronization. Nodes  exchange data, verify transactions, and reach consensus on the state of the blockchain. 
  \item \emph{Consensus layer.} This layer employs consensus protocols to verify the legitimacy of transactions before they are added to the blockchain, and ensures a consistent view of the blockchain's history for all nodes. 

  \item \emph{Contract layer.} Smart contracts facilitate automated execution of predefined actions, such as releasing relief funds when specific criteria are met or triggering delivery operations upon confirmation of demand. Thereby, it fosters well-organized and collaborative disaster response.

  \item \emph{Service layer.} 
  By leveraging the transparency of blockchain, the service layer enables real-time tracking of relief supplies, donations, and their utilization in various UDRN services, fostering accountability and trust.
\end{itemize}

\vspace{-3mm}
\subsection{Key Demands of Blockchain-Envisioned UDRNs}
\begin{itemize}
    \item \textit{Low latency.} In blockchain-empowered UDRNs, timely disaster information delivery and low transaction latency are critical for rapid disaster response and effectively coordinated relief distribution.
    \item \textit{High scalability.} In large-scale UDRNs, the number of participating entities (e.g., UAVs, rescue vehicles, and GSs) and data transactions can be substantial. Ensuring blockchain scalability is crucial to handle increased network traffic and maintain efficient relief operations.
    \item \textit{Strong security.} Security is paramount to maintain integrity, trustworthiness, and traceability of relief operations in blockchain-based UDRNs, by preventing relief contract vulnerabilities and resisting transaction malleability threats in a real-time and energy-efficient manner.
\end{itemize}

\vspace{-4mm}
\subsection{State-of-the-art Blockchain Approaches for UDRNs}
In the literature, various works propose blockchain-based approaches to secure and optimize UDRNs. Su \emph{et al.} design a lightweight blockchain named \emph{LVBS}\cite{lvbs} {among rescue vehicles and UAVs} to secure collaborative air-ground networking for disaster data sharing, {where a credit-based consensus protocol is devised to trace entities' misbehaviors.} Wang \emph{et al.} develop {a partition-tolerant and energy-efficient} blockchain named \emph{RescueChain} \cite{rescuechain} along with a reputation-based Tendermint consensus protocol, to safeguard data sharing in post-disaster sites. Xing \emph{et al.} \cite{XingTVT2022} design a delay-tolerant blockchain system aided by UAVs for secure offline transactions, {where offline payment channels are established by hashed time locked contracts to resist deposit forgery.} Wang \emph{et al.} \cite{WangTNSM2024} design an infrastructure-free and lightweight consortium blockchain system for disaster rescue in UAV-assisted Internet of vehicles (IoV) through threshold signature, pre-selection and group scoring mechanisms. 
Pauu \emph{et al.} \cite{PauuIoTJ2023} present a blockchain-enabled UAV-assisted decentralized graph federated learning (GFL) scheme for secure knowledge sharing in disasters, {where blockchain ensures the integrity of model weights in GFL.} Kaur \emph{et al.} \cite{10004889} develop an Ethereum blockchain-based decentralized donation mechanism for transparent charity donations {under emergencies}, where smart contracts {including registration contract, beneficiary contract, and donor contract} are deployed to automatically process donations. 

In the following, we discuss the key challenges in existing blockchain approaches for UDRNs in Sect.~\ref{challenges} and present the potential solutions to resolve them in Sect.~\ref{solutions}.

\vspace{-1mm}
\section{Challenges of Blockchain-Envisioned UDRNs}\label{challenges}
This subsection highlights the key challenges towards secure and efficient blockchain-envisioned UDRNs.

\vspace{-2mm}
\subsection{Lack of Cooperation Between Smart Contracts in UDRNs}\label{Challenge1}


Smart contracts enable automated transactions and operations within UDRNs, eliminating the need for intermediaries. 
However, entities involved in disaster relief management, such as government agencies, NPOs, and local community groups, often have distinct and even competitive priorities and objectives. Consequently, smart contracts designed for each entity often operate in silos \cite{10004889}, leading to fragmented and duplicated response efforts. For instance, independent smart contracts might execute redundant or conflicting actions, such as sending multiple UAVs to the same location when fewer would suffice, resulting in wasted time and resources. Avoiding resource misallocation is crucial in relief management. Without proper coordination between smart contracts, relief resources such as medical supplies, food, and water might be dispatched to areas that do not require them urgently, while critical regions remain underserved. Additionally, timely and accurate information is crucial during disaster relief. A lack of shared data or inconsistent data updates between smart contracts can lead to misinformed decisions, causing delays in relief operations. Therefore, a well-coordinated approach for smart contracts is necessary to ensure seamless relief distribution, efficient resource allocation, and timely assistance.

\vspace{-2mm}
\subsection{Lack of Dynamic Audit for Smart Contract Vulnerabilities in UDRNs}\label{Challenge2}
Smart contracts on UDRNs, while ensuring immutability and trust-free relief management, present significant security challenges due to their immutable nature once deployed. This rigidity means that any vulnerability in contract codes after deployment cannot be rectified or patched. Consequently, smart contracts are susceptible to a variety of vulnerabilities, including unchecked call attack, timestamp dependency attack, and reentrancy attack. Traditional static vulnerability audit methods \cite{securify} \cite{oyente} are limited to pre-deployment audits and lack dynamic real-time vulnerability audit capabilities in post-deployment stage. Although existing dynamic vulnerability audit methods \cite{evil-under-the-sun}\cite{TXSPECTOR} work in post-deployment stage, they suffer from high execution costs, limited EVM bytecode size, and difficulty in covering all smart contracts.
{Moreover, in UDRNs, the logic dependence of collaborative relief contracts complicates the detection of security vulnerabilities.
Additionally,} UDRNs operate in environments that are constantly changing, with variables such as disaster conditions, resource availability, and frequently fluctuating participants. 
Real-time auditing enables continuous monitoring and prompt response to suspicious activities in UDRNs.
Hence, it necessitates a dynamic audit approach that quickly adapts to new threats and conditions while offering real-time detection for evolving contract vulnerabilities in UDRNs.

\vspace{-2mm}
\subsection{Low Forensics Robustness under Transaction Malleability Attacks in UDRNs}\label{Challenge3}
Transactions in UDRNs play a crucial role in recording the ledger of relief services. They ensure that all actions and resource allocations are transparently documented on the blockchain, providing an accountable and traceable forensic history of disaster relief efforts. However, transaction malleability attacks \cite{Trans_Unmalleable}, including transaction manipulation and data tampering, can jeopardize the integrity of the forensic history and facilitate fraudulent activities during relief management. These attacks can alter transactions before they are confirmed on the blockchain and cause inconsistencies in transaction records, making it challenging to verify their authenticity and integrity. {This is particularly severe in UDRN scenarios, where timely and accurate information is crucial for effective relief efforts under cooperative relief contracts.} Additionally, transaction forensics depends on the ability to trace and audit transactions accurately, where the lack of accountability can lead to mistrust in UDRNs and hinder collaborative efforts during relief operations.
Hence, robust transaction forensics mechanisms are essential to defend against transaction malleability attacks in blockchain-based UDRNs, ensuring the reliability and trustworthiness of the disaster relief process.


\section{Solutions to Blockchain-Envisioned UDRNs}\label{solutions}

To resolve the above key challenges, this section {first devises} a series of collaborative smart contracts for automated and coordinated relief management in UDRNs (in Sect.~\ref{Opportunities1}). {Under collaborative contract environments with high dependency, we further propose} a dynamic contract audit mechanism (in Sect.~\ref{Opportunities2}) {to prevent contract vulnerabilities} and a robust transaction forensics strategy (in Sect.~\ref{Opportunities3}) {to resist transaction malleability attacks.}

\begin{figure}[t]\setlength{\abovecaptionskip}{-0.03cm}
  \centering
  \includegraphics[width=1.02\linewidth]{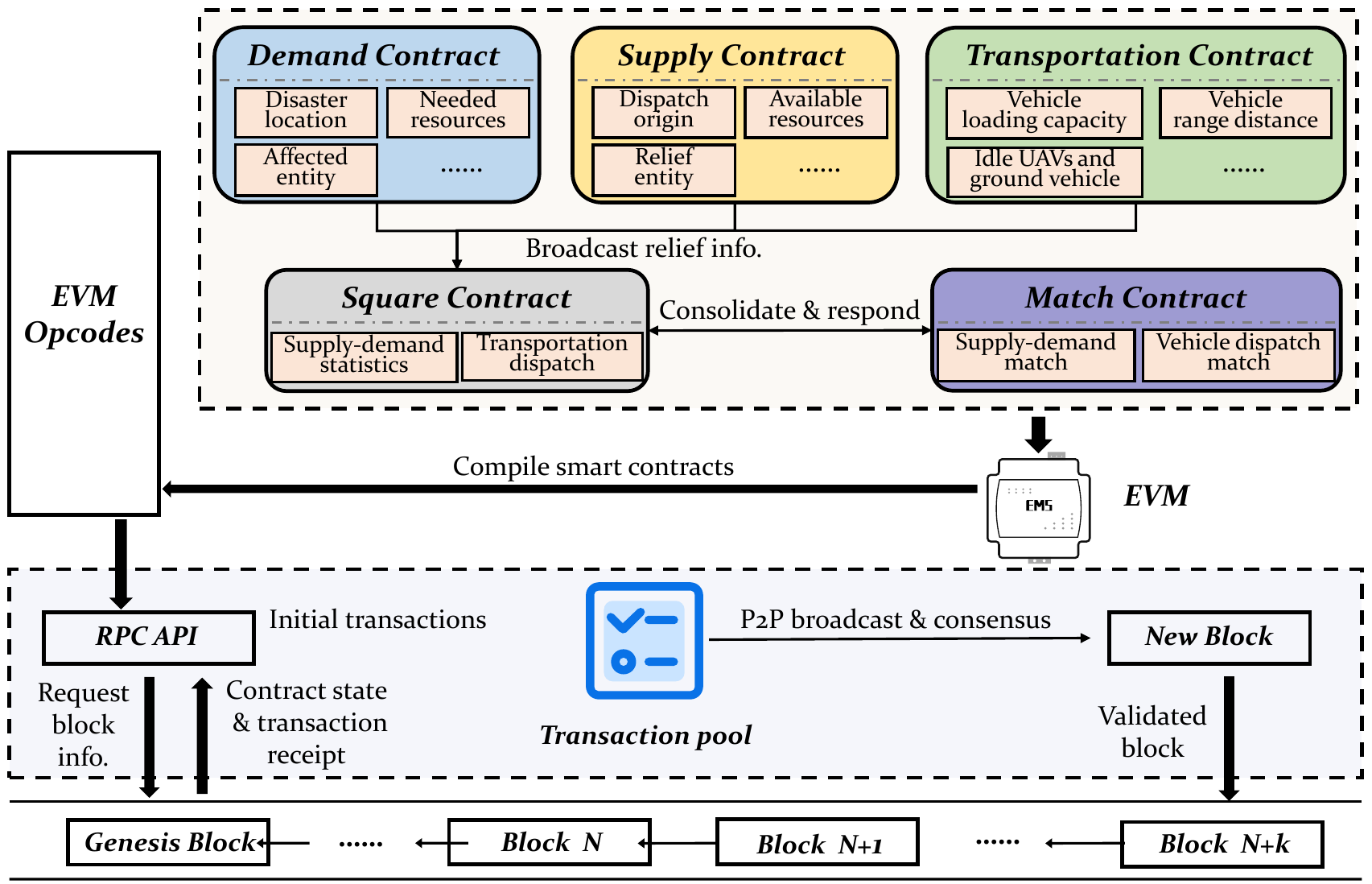}
    \caption{Illustration of collaborative smart contracts design for automated and coordinated disaster relief management in UDRNs.}
  \label{smart_contract}\vspace{-3mm}
\end{figure}

\subsection{Collaborative Smart Contracts for Relief Management in UDRNs}\label{Opportunities1}
As shown in Fig.~\ref{smart_contract}, we have designed five types of smart contracts which operate cooperatively to enable coordinated and automated relief management in UDRNs.
These contracts are encapsulated as program codes and deployed on the Ethereum virtual machine (EVM) through the following six steps. \ding{172}The contracts are compiled into bytecode. \ding{173}Contract creation transactions are generated using remote procedure call (RPC). \ding{174}Contract transactions are validated and subsequently added to the transaction pool. \ding{175}Miners package new transactions from the the transaction pool, generate a new block, and broadcast it across the entire blockchain network.
\ding{176}Consensus nodes run the consensus protocol to reach agreement on the new block to be appended to the blockchain. \ding{177}Finally, addresses of these contracts and receipt for these transactions are generated, indicating the successful deployment of contracts.

Specifically, five types of smart contracts are defined.
\begin{itemize}
\item \emph{Demand Contract:}
It specifies the process for requesting and receiving disaster relief materials. The affected population can invoke this contract by providing demand-related information, including location, types, and quantities of needed relief resources. {It enables direct reach to disaster-affected individuals through public APIs and functions that organize and prioritize requests based on urgency and availability.}
    \item \emph{Supply Contract.}
It specifies the donation and response process of disaster relief materials. Individuals, government agencies, NPOs, and local community groups can invoke this contract by providing supply-related information, including location, types, and quantities of denoted relief resources. {It streamlines the donation process and ensures accurate tracking of available supplies.}
    \item \emph{Transportation Contract.}
It specifies the process of collaborative resource transportation. Idle UAVs and ground vehicles can invoke this contract by sending transportation-related information, including payload capacity and driving/flying range. 
    \item \emph{Square Contract.}
It aggregates all instances of demand contracts, supply contracts, and transportation contracts. GSs and the relief cloud can invoke this contract to view relief supply-demand statistics and transportation statuses of relief materials {for strategic resource allocation decisions.}
    \item \emph{Match Contract.}
It specifies the process of relief supply-demand matching and UAV/vehicle dispatch matching. First, it queries all demand and supply contract instances in the square contract and performs matches based on resource types and quantities. Then, for all matched resources, it calculates transportation distance and efficiency requirements. Subsequently, based on the transportation contract instances in the square contract, it allocates UAVs and vehicles with the corresponding transportation capabilities to collaboratively complete transportation assignments of relief materials. 
\end{itemize}

The above five types of smart contracts collaborate and interact for coordinated relief management as follows. Initially, affected population and relief organizations employ demand contracts and supply contracts to submit their resource demands and supplies, respectively. Then, UAVs and vehicles broadcast their transportation capabilities via transportation contracts. Next, the square contract consolidates all on-chain relief information, while the matching contract facilitates relief supply-demand matching and arranges transportation dispatches.

\begin{figure}[t]\setlength{\abovecaptionskip}{-0.03cm}
  \centering
  \includegraphics[width=1.02\linewidth]{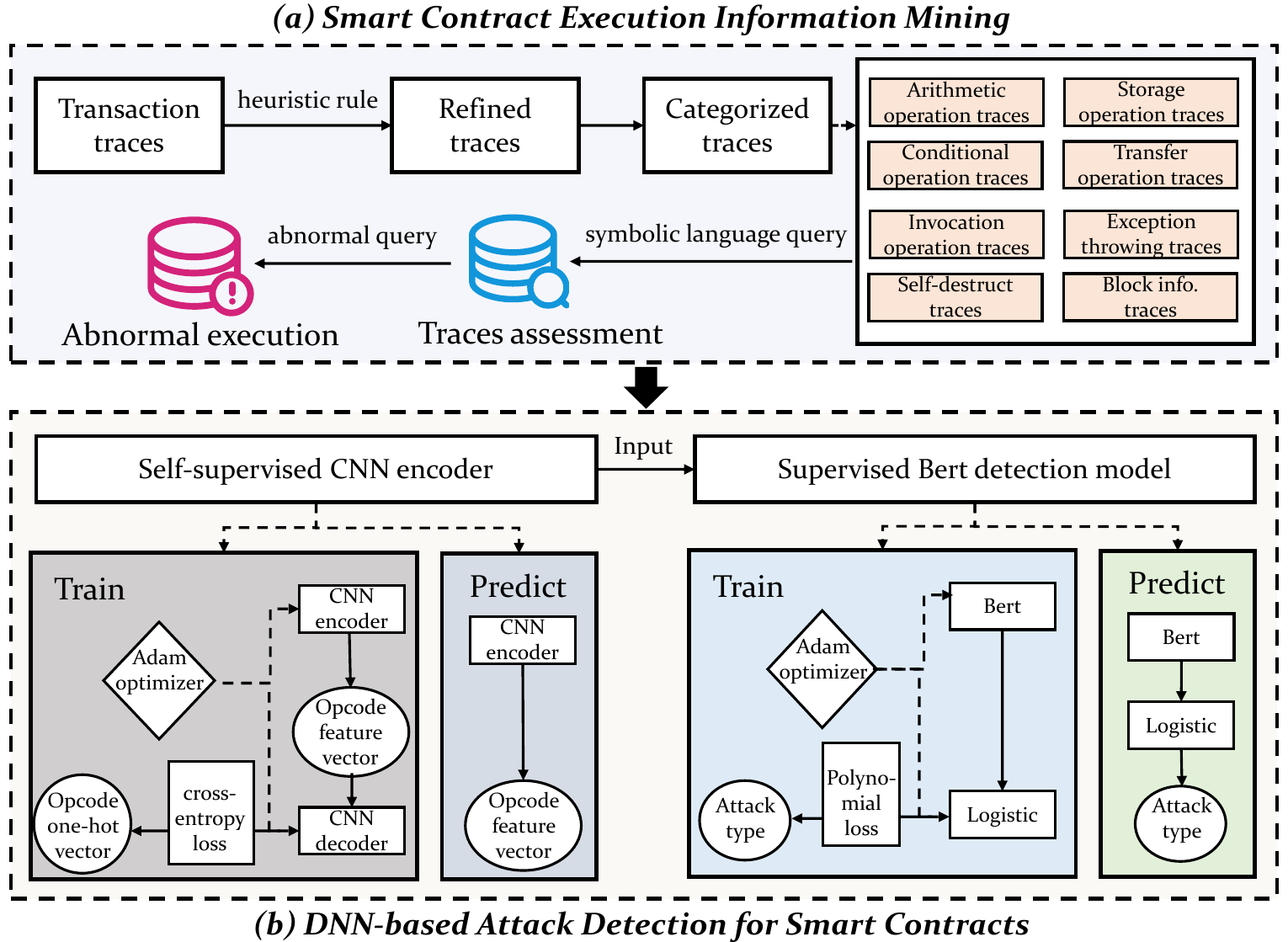}
    \caption{Illustration of AI-based dynamic smart contract vulnerabilities audit including: (a) smart contract execution information mining, and (b) DNN-based vulnerability detection for smart contracts.}
  \label{contract_audit}\vspace{-3mm}
\end{figure}

\vspace{-3mm}
\subsection{Dynamic Smart Contract Vulnerabilities Audit in UDRNs}\label{Opportunities2}
This subsection devises (i) a contract execution data mining mechanism to uncover potential risks through real-time underlying data collection of smart contract operations; and (ii) an intelligent attack detection model based on deep neural networks (DNN) to dynamically assess potential contract-related threats.

\emph{1) Low-cost smart contract execution information mining.} As depicted in Fig.~\ref{contract_audit}, it includes the following 4 steps.
\ding{172}First, we utilize the Geth tool to obtain a complete transaction trace of contracts, including EVM-executed opcodes, program counters, call stack depth, current stack values, etc.
\ding{173}By pinpointing the contract address, we dynamically synchronize and mine contract execution information (i.e., transactions) by real-time tracking and replaying any transaction initiated or received at the contract address.
\ding{174}Furthermore, we devise heuristic rules to remove irrelevant transaction traces from the transaction tracker, such as opcode program counters and gas consumption. As such, the mining cost of contract information can be reduced.
\ding{175}Finally, we categorize the mined contract execution information into eight types, based on the functionality of EVM instructions: traces for arithmetic operations, storage operations, conditional operations, transfer operations, invocation operations, exception throwing, self-destruct actions, and block information. Using the categorized information and symbolic language queries, we determine the contract's operational status and list potential abnormal transaction executions for further detection and assessment.

\emph{2) DNN-based attack detection for smart contracts.}
The opcode sequence, representing the operational logic of a smart contract, is crucial for assessing potential contract vulnerabilities by analyzing abnormal transaction executions. As shown in Fig.~\ref{contract_audit}, we first use a self-supervised convolutional neural network (CNN) {with 5 layers of convolutional encoder/decoder} to encode the opcode sequences of abnormal transaction executions to capture their features. Next, a supervised BERT model is trained to identify and classify the features of abnormal transaction executions, determining whether these abnormal transactions constitute vulnerabilities against smart contracts. If so, we further classify the vulnerability type. The BERT model for abnormal transaction classification is deployed on the cloud for online updating and training. This allows continuous augmentation of its knowledge base with new attack categories to dynamically expand its vulnerability database. Additionally, newly identified attack types, audited by experts, can be self-updated to the cloud, thereby enabling adaptive detection of previously unknown types of contract vulnerabilities.

\vspace{-3mm}
\subsection{Robust Transaction Forensics with On/off-chain Cooperation}\label{Opportunities3}

To resist transaction malleability attacks \cite{Trans_Unmalleable} in UDRNs, a robust transaction forensics mechanism is devised based on smart contracts, which consists of the following two phases.

\begin{figure}[!t]\setlength{\abovecaptionskip}{-0.03cm}
  \centering
  \resizebox{1.05\linewidth}{!}{\includegraphics{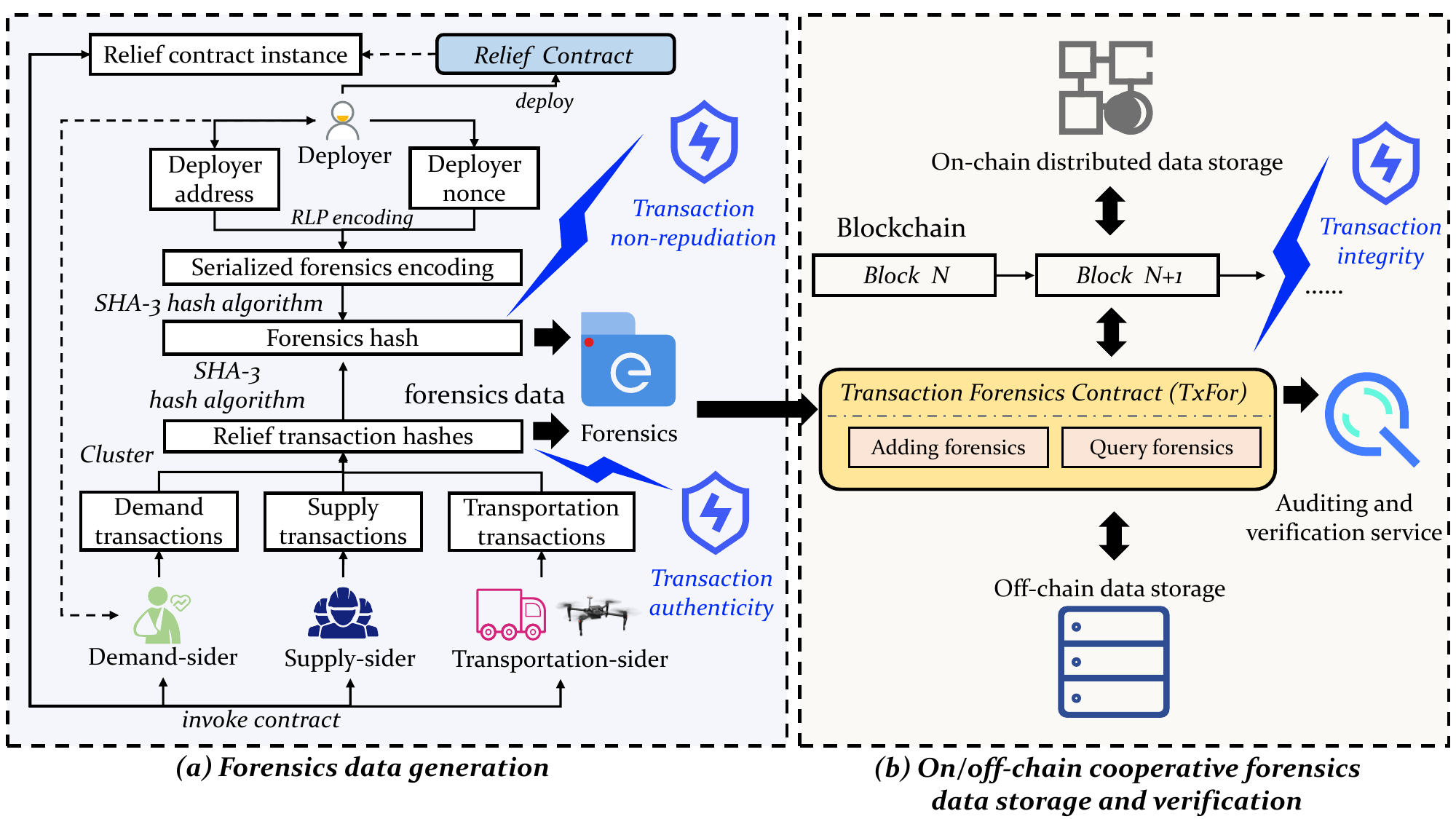}}
  \caption{Illustration of robust transaction forensics services in UDRNs including: (a) trusted forensics data generation, and (b) on/off-chain cooperative forensics data storage and verification.}
\label{trans_forensics}\vspace{-3mm}
\end{figure}

\emph{1) Trusted forensics data generation.} As depicted in Fig.~\ref{trans_forensics}, the forensics data comprises two components: forensics hash and forensics content (i.e., relief transaction hashes). Relief transaction hashes are generated when the demand-side, supply-side, and transportation-side invoke their respective smart contracts, and are then recorded and clustered based on the contract instances they interacted with. Consequently, forensics content activated within the same relief contract instance are associated with a distinct forensics hash, {enabling authenticity verification of relief transactions and holding entities accountable for their actions.} 

For the forensics hash, it is generated during the contract deployment phase through two steps: \ding{172}Recursive length prefix (RLP) serialization encoding is applied using the nonce and the blockchain address of the contract deployer. The nonce increments each time a contract is deployed, {ensuring the uniqueness and non-repudiation of the serialized encoding for each contract instance.} \ding{173}The serialized encoding, along with the transaction hashes invoked by the same entity, are input into the Secure Hash Algorithm 3 (SHA-3) to produce the final forensics hash.
Since contract deployers in UDRNs are the same entities that invoke their respective smart contracts, the forensics hash serves as evidence for authenticating the actions of contractual entities in the forensics process, enabling non-repudiation of their corresponding relief transactions.

\emph{2) On/off-chain cooperative forensics data storage and verification.} We design a transaction forensics (TxFor) contract for distributed storage and management of forensics data, {ensuring the integrity of relief transactions.} Here, forensics data is stored in an on/off-chain collaborative manner, with data access rules defined through smart contracts. We maintain a map dictionary to link each forensics hash to its corresponding forensics content. Specifically, forensics contents including relief transaction hashes are stored in the off-chain data store, while the corresponding hash pointers are maintained on distributed ledgers, to alleviate on-chain data storage and synchronization costs. This on/off-chain collaborative method facilitates the implementation of functions to add and query transaction forensics on the blockchain. By conducting high-frequency data interactions off-chain while reserving on-chain activities for auditing, verification, and other safety-critical operations, the service response latency in blockchain-based UDRN forensics services can be reduced.

\section{Implementation and Evaluation}
We implement a prototype of the proposed blockchain system on a server running Ubuntu 20.04 OS, equipped with an Intel Xeon Gold 6271C CPU, 256GB memory, and dual NVIDIA RTX 3090 graphics cards.
For system implementation, Geth (v1.7.0) is used to establish a blockchain test network. Truffle suite (v4.1.12) is employed to compile, test, and deploy smart contracts, all of which are written in Solidity (v0.5.13). We utilize Web3.js (v1.2.6) to design a dynamic verification interface, which calls txFor contract to dynamically retrieve forensics data from the blockchain and provide verification services. We employ Solc.js (v0.5.13) as the smart contract compilation tool. All the proposed three solutions are implemented in the prototype blockchain network. Our prototype is tested to support 312 full nodes and 2,184 light nodes.
The CNN-based auto-encoder is trained via Mean Squared Error (MSE) loss, while the BERT model is trained via cross-entropy loss. Both of them are trained via Adam optimizer with learning rate of 0.01 and dropout.

\begin{figure}[ht]
\fbox{\centering\setlength{\abovecaptionskip}{-0.1cm}
\includegraphics[width=6.8cm]{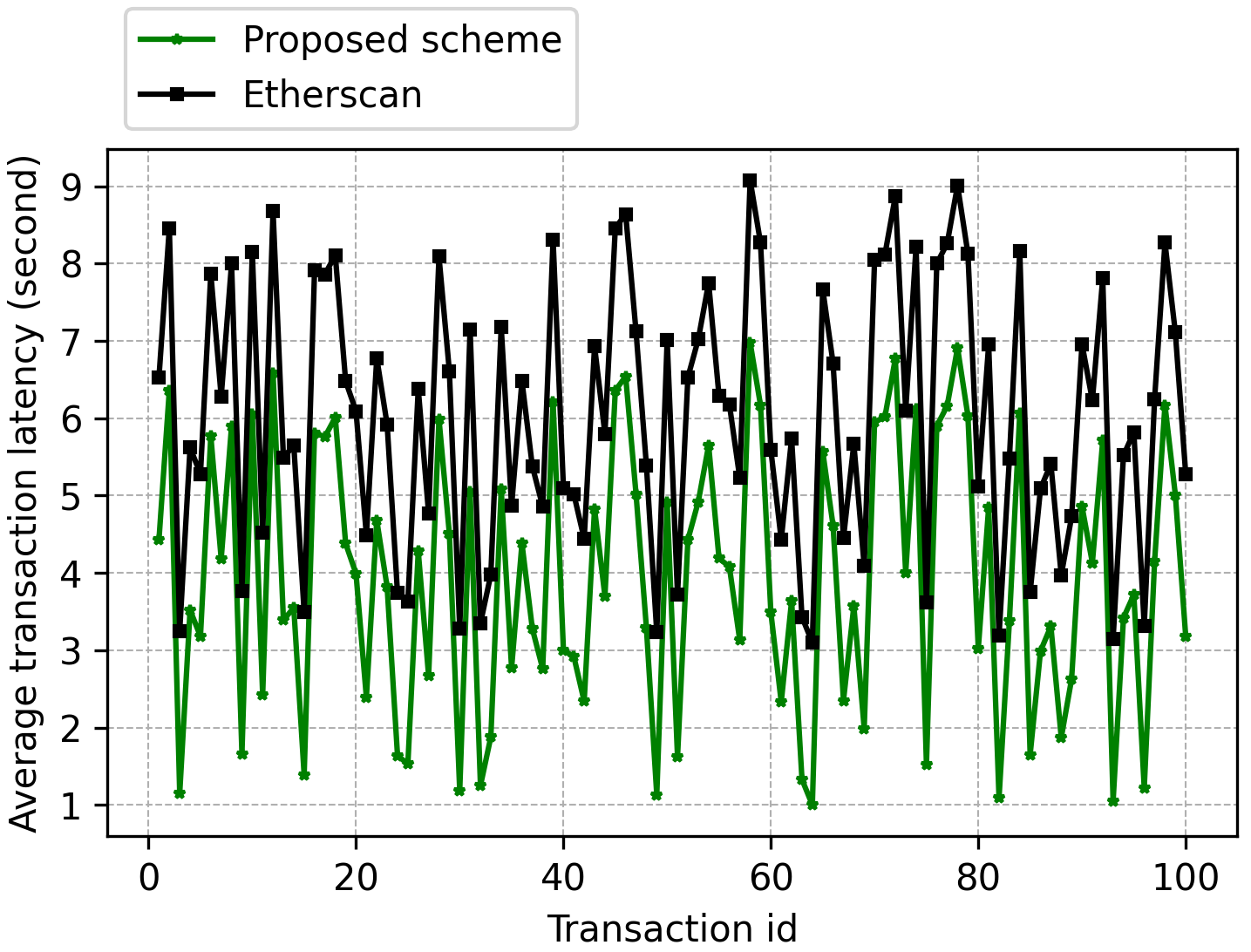}}
\caption{Comparison of average transaction latency in the proposed scheme and Etherscan.}
\label{evaluation:mining}\vspace{-2.5mm}
\end{figure}

\begin{figure}[ht]
\fbox{\centering\setlength{\abovecaptionskip}{-0.1cm}
\includegraphics[width=6.8cm]{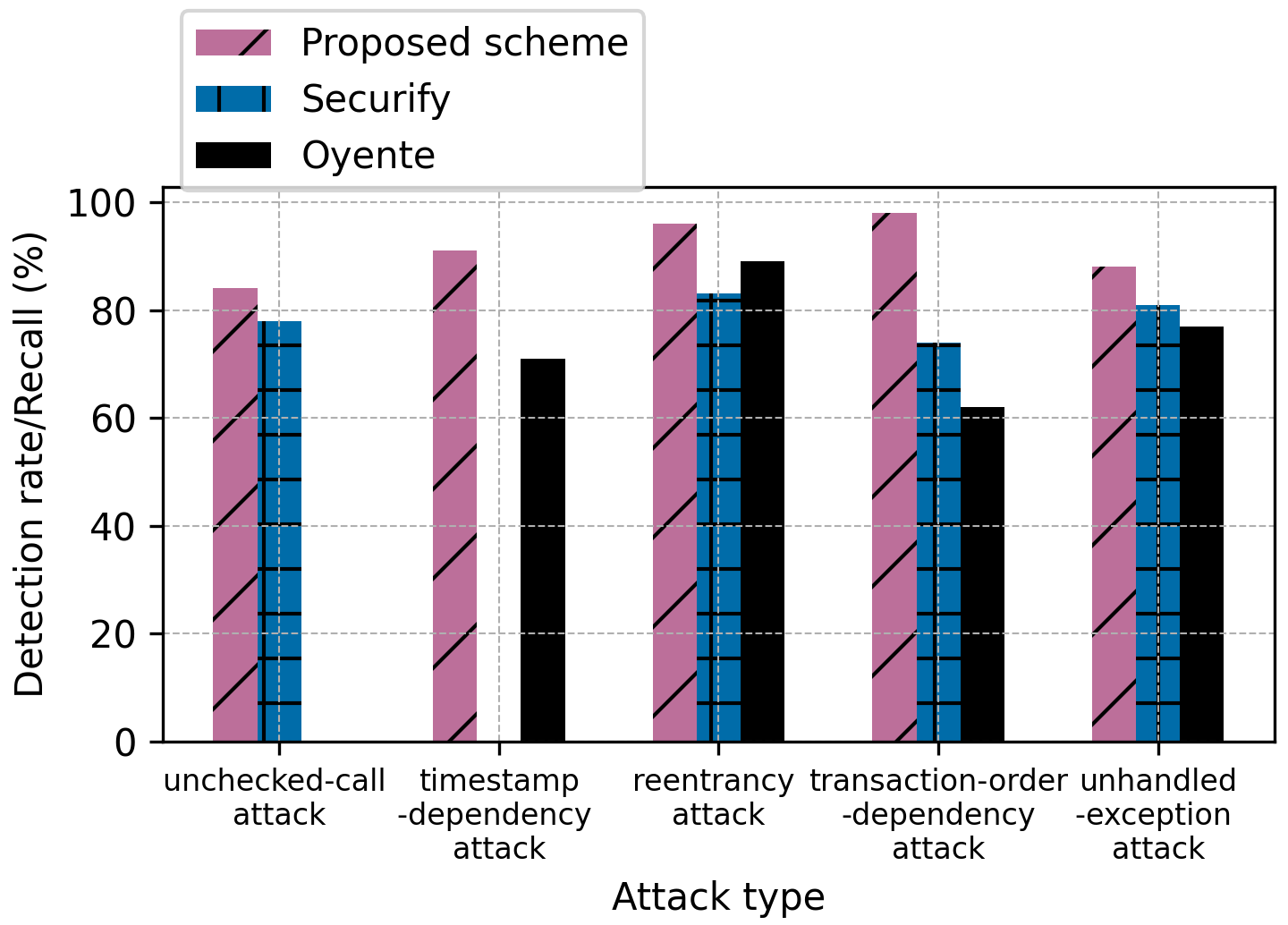}}
\caption{Comparison of detection rate in the proposed scheme, Securify \cite{securify}, and Oyente \cite{oyente}, under five typical smart contract vulnerabilities.}
\label{evaluation:detection}\vspace{-3mm}
\end{figure}


As shown in Fig.~\ref{evaluation:mining}, Etherscan (i.e., a well-known Ethereum blockchain online analysis platform) incurs an average transaction latency of 6.04 seconds, with a maximum latency of up to 9.08 seconds per transaction. In contrast, our prototype reduces the average transaction latency to 3.94 seconds and the maximum latency to 6.98 seconds per transaction, thereby significantly enhancing the efficiency of disaster relief contracts. {Here, the transaction latency consists of three parts: (i) transaction confirmation time in collaborative contracts, (ii) robust transaction forensics time, (iii) mining time of contract execution information.} 

Next, we compare the proposed scheme with two mainstream contract vulnerability detection schemes: Securify \cite{securify} and Oyente \cite{oyente}. 
Five typical attacks on smart contracts are reproduced: reentrancy, unchecked call, timestamp dependency, transaction order dependency, and unhandled exception attacks.
As depicted in Fig.~\ref{evaluation:detection}, Oyente fails to detect unchecked call attack while Securify fails to detect timestamp dependency attack. In contrast, the proposed scheme effectively detects all five types of vulnerabilities and achieves the highest detection rate.
Compared to the best-performing baseline, the proposed scheme achieves the following improvements in detection rate (i.e., recall, a crucial metric in vulnerability detection, measuring the completeness of positive predictions): a 3.1\% increase in reentrancy attacks, a 5.3\% increase in unchecked call attacks, a 24.7\% increase in timestamp dependency attacks, a 28.4\% increase in transaction order dependency attacks, and a 4.6\% increase in unhandled exception attacks. 

\section{Open Research Issues}

\subsection{Generative AI (GAI) and Large Language Models (LLMs) for Securing and Optimizing Blockchains in UDRNs}
{GAI, especially LLMs, can revolutionize future blockchain design in UDRNs by enhancing security, privacy, scalability, and interoperability. Nguyen \textit{et al.} \cite{GAI_Nguyen} review the utilization of GAI approaches in optimizing and securing blockchains, e.g., blockchain configuration optimization and smart contract vulnerability detection. They also devise a generative diffusion model-based approach to fine-tune block producer selection, block time, and block size within a consortium blockchain for Internet of things. Gai \textit{et al.} \cite{LLM_Gai} develop BLOCKGPT, a LLM-based Ethereum transaction anomaly detection tool that dynamically identifies suspicious or malicious on-chain activities.
However, LLMs may also introduce additional vulnerabilities such as hallucination, prompt injection, and data memorization risks in blockchain-empowered UDRNs.
Additionally, leveraging LLMs to dispatch UAVs and establish resilient communications for enhanced connectivity in the blockchain system across space, air, and ground platforms under disasters remains an open issue.}

\subsection{Semantic Communications for Optimized UDRNs}
By focusing on the meaning of transmitted information rather than raw data, semantic communications can prioritize critical messages and reduce bandwidth usage in resource-constrained and time-varying UDRNs. This allows satellites, UAVs, and ground robots to understand and prioritize essential rescue-related information, enabling context-aware disaster data exchange. However, ensuring robust and accurate semantic interpretation across space-air-ground disaster rescue systems is a significant challenge. Additionally, maintaining the reliability and security of semantic data transmission in dynamic disaster environments remains an ongoing concern.

\subsection{Low-Cost \& Scalable Cross-Chain Mechanisms in Disaster}
In practical disaster relief applications, diverse stakeholders may deploy relief services based on distinct blockchains, due to their different technological infrastructures and service requirements. Cross-chain technology enables interoperability between different blockchain networks. In UDRNs, optimizing energy consumption and enhancing scalability are critical goals for developing robust and cost-effective cross-chain mechanisms, especially given network conditions such as intermittent connectivity and limited bandwidth.

\section{Conclusion}
In this article, we have discussed the technical challenges and potential solutions in blockchain-enabled URDNs to promote secure and efficient disaster response and relief management. We first introduced a general architecture of blockchain-enabled UDRNs. Then, we examined state-of-the-art approaches, identified three key efficiency and security challenges, and devised corresponding solutions for blockchain-enabled UDRNs. A real prototype was implemented for evaluation. We also discussed several open research issues. By harnessing the potential of blockchain technology in UDRNs, this study aims to contribute to more resilient and collaborative disaster response efforts, ultimately saving lives in critical disaster situations.


\end{document}